\documentclass{ieeeaccess}
\usepackage{cite}
\usepackage{amsmath,amssymb,amsfonts}
\usepackage{graphicx}
\usepackage{textcomp}
\usepackage{bbm}
\usepackage{caption}
\usepackage{hyperref}
\usepackage{algpseudocode}

\usepackage{algorithm} 
\usepackage{algorithmicx}

\def\BibTeX{{\rm B\kern-.05em{\sc i\kern-.025em b}\kern-.08em
    T\kern-.1667em\lower.7ex\hbox{E}\kern-.125emX}}
\begin{document}
\doi{10.1109/TQE.2020.DOI}

\title{MIMO with $1$-bit Pre/Post-Coding Resolution: A Quantum Annealing Approach}
\author{\uppercase{Ioannis Krikidis}\authorrefmark{1} \IEEEmembership{Fellow, IEEE}}
\address[1]{Department of Electrical and Computer Engineering, University of Cyprus, Cyprus (email: krikidis@ucy.ac.cy)}
\tfootnote{This work received funding from the European Research Council (ERC) under the European Union’s Horizon 2020 research and innovation programme (Grant agreement No. 819819) and the European Union’s Horizon Europe programme (ERC PoC, Grant agreement No. 101112697). It has also received funding from the European Union HORIZON programme under iSEE-6G GA No. 101139291.}

\markboth
{I. Krikidis : MIMO with $1$-bit Pre/Post-Coding Resolution: A Quantum Annealing Approach}
{I. Krikidis: MIMO with $1$-bit Pre/Post-Coding Resolution: A Quantum Annealing Approach}

\corresp{Corresponding author: Ioannis Krikidis (email: krikidis@ucy.ac.cy).}

\begin{abstract}
In this paper, we study the problem of digital pre/post-coding design in multiple-input multiple-output (MIMO) systems with $1$-bit resolution per complex dimension. The optimal solution that maximizes the received signal-to-noise ratio relies on an NP-hard combinatorial problem that requires  exhaustive searching with exponential complexity. By using the principles of alternating optimization and quantum annealing (QA), an iterative QA-based algorithm is proposed that achieves near-optimal performance with polynomial complexity. The algorithm is associated with a rigorous mathematical framework that  casts the pre/post-coding vector design to appropriate real-valued quadratic unconstrained binary optimization (QUBO) problems. Experimental results in a state-of-the-art D-WAVE QA device validate the efficiency of the proposed algorithm. To further improve the efficiency of the D-WAVE quantum device, a new pre-processing technique which preserves the quadratic QUBO matrix from the detrimental effects of the Hamiltonian noise through non-linear companding, is proposed. The proposed pre-processing technique significantly improves the quality of the D-WAVE solutions as well as the occurrence probability of the optimal solution. 
\end{abstract}


\titlepgskip=-15pt

\maketitle

\section{Introduction}

\PARstart{A}{fundamental} enabling technology for future 6G wireless communication systems is massive/large multiple-input multiple-output (MIMO) systems \cite{ZHE}. By exploiting the spatial degrees of freedom associated with the antenna arrays and through appropriate signal processing, MIMO systems are able to support the extremely high requirements of capacity, throughput and reliability. However, as the number of antennas increases, the implementation complexity/cost of conventional digital MIMO in terms of radio frequency (RF) chains, digital-to-analogue converters (DACs) and/or analogue-to-digital converters (ADCs), base-band signal processing {\it etc.} becomes a bottleneck for practical applications \cite{ALI}. To address this challenge, analogue/digital hybrid MIMO architectures which split the processing into digital and analogue domain to balance the trade-off between performance and implementation complexity/cost  have emerged as a promising solution \cite{ALI}.  Other relevant techniques consider analogue-only signal processing where low-resolutions phase shifters are used to implement analogue pre/post-coding (mainly suitable for high frequency bands {\it e.g.,} millimeter wave), or the employment of low-resolution DACs/ADCs \cite{SIL,LOZ}. 

In contrast to existing approaches, in this work, we study a basic (digital) point-to-point MIMO system with low-resolution pre/post-coding processing; specifically, we assume that both the transmitter and receiver employ $1$-bit signal processing resolution ($1$-bit per complex dimension) in order to extremely reduce complexity and power consumption\footnote{It is worth noting that although $1$-bit processing at the transmitter is similar to $1$-bit DAC, the assumption of $1$-bit processing at the receiver is fundamentally different to conventional $1$-bit ADC; this assumption, makes the considered architecture different from the existing $1$-bit DAC/ADC MIMO architectures \cite{LOZ}.}. This new MIMO architecture is promising for future low-power/low-computation devices in the era of the Internet of Things. Due to the complex-valued binary resolution of the new MIMO architecture, the design of pre/post-coding vectors becomes an NP-hard problem; it can be solved optimally through exhaustive search (ES) over all the possible transmit/receive configurations corresponding to an exponential complexity. To the best of the author's knowledge, the considered MIMO design problem is reported for the first time in this paper.   

Adiabatic quantum computing is a promising tool to solve NP-hard problems by using the principles of the Adiabatic Theorem \cite{MCG}. Specifically, if a quantum system is initially in the ground state (quantum state with the lowest energy) of an initial Hamiltonian and the system evolves/changes slowly (gradually), it will converge to the ground state of the final Hamiltonian. By encoding the final Hamiltonian to the desired problem, the Adiabatic evolution can be used to solve complex combinatorial optimization problems which are represented as instances of the Ising model (equivalent to quadratic unconstrained binary optimization (QUBO) problem) \cite{KAS}. D-WAVE is a commercial analogue quantum device that implements a noisy approximation of the quantum Adiabatic algorithm called quantum annealing (QA) \cite{MCG,WAVE}. This quantum device has received considerable interest lately due to the high number of available qubits (more than $5,000$ qubits in latest hardware architectures) and its friendly interface for remote access. It is worth noting that in QA, instead of analyzing the computation time/complexity of a given algorithm, we mainly study the trade-off between time and the probability that the output is correct. 

D-WAVE QA is introduced as a promising computation tool to solve complex (NP-hard) combinatorial problems in wireless communication system \cite{KIM}. The associated literature mainly focuses on the solution of the maximum-likelihood detection problem in large multi-user MIMO systems \cite{JAM1,JAM2}; recent research studies employ D-WAVE QA in other wireless communication problems such as channel decoding (polar codes) \cite{KASI}, passive beamforming design in intelligent reflecting surfaces \cite{ROSI}, configuration selection in reconfigurable/fluid-antenna systems \cite{KRI}, beam assignment in satellite systems \cite{DIN} {\it etc.}. The integration of D-WAVE QA in the design of wireless communication systems is a new research area of paramount importance. 

In this paper, we extend our work in \cite{KRI3} towards the design of $1$-bit digital (complex-valued) pre/post coding vectors. Specifically, we solve the problem of pre/post-coding design in a MIMO system with $1$-bit resolution per complex dimension. We focus on the maximization of the received signal-to-noise ratio (SNR) which due to the complex binary assumption relies to the solution of an NP-hard complex-valued combinatorial problem. By using the principles of alternating optimization and the computation capabilities of the D-WAVE QA, an iterative algorithm is proposed which solves appropriate QUBO instances at each iteration to gradually compute the pre/post-coding vectors. An appropriate mathematical framework that converts the original complex-valued combinatorial problem to real-valued QUBO instances is investigated. Experimental results in a state-of-the-art D-WAVE QA device show that the proposed iterative algorithm achieves near-optimal performance (similar to ES) while ensuring polynomial complexity.      

To further improve the quality of the returned distinct D-WAVE QA solutions, a new pre-processing technique that preserves the quadratic QUBO matrix from the detrimental effects of the Hamiltonian noise is proposed. The proposed pre-processing technique is motivated by the non-uniform compression in pulse code modulation (PCM) communication systems. Specifically, motivated by the experimental observation that small magnitude elements ($\approx 0$) of the quadratic matrix are more sensitive to the Hamiltonian noise \cite{ICE} (similar to quantization noise in PCM \cite{PRO}), a non-linear {\it companding} ($\mu$-law) pre-processing is proposed to boost small-magnitude elements without qualitatively affecting the structure of the original QUBO problem. Experimental results show that the proposed pre-processing technique improves the quality of the D-WAVE solutions (we have less distinct solutions of higher quality) and increases the probability of the optimal solution.   

The remainder of this paper is organized as follows. The system model and the problem formulation are described in Section \ref{sec1}. Section \ref{sec2} presents the proposed QA-based iterative algorithm for the design of the $1$-bit pre/post-coding vectors. The QUBO pre-processing technique that improves the efficiency of the D-WAVE QA is presented in Section \ref{sec3}. Pertinent experimental examples are presented in Section \ref{sec4}. Finally, concluding remarks are discussed in the last section. 

\begin{figure}
\centering
\includegraphics[width=\linewidth]{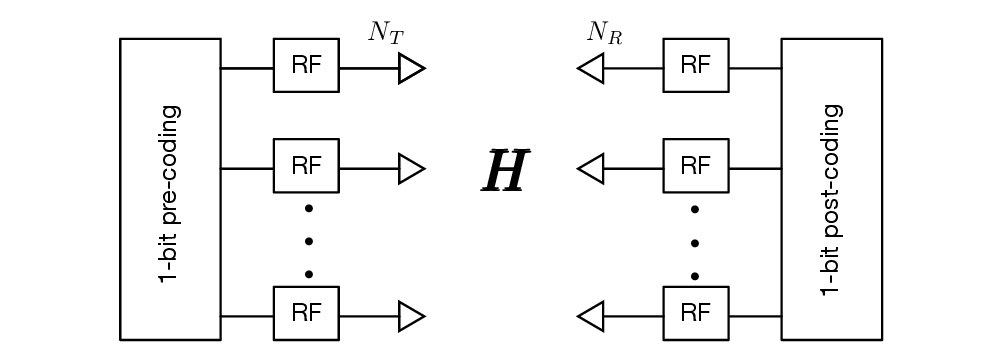}
\caption{$N_R\times N_T$ MIMO channel with $1$-bit pre/post-coding vectors.}\label{fig_sys}
\end{figure}

{\it Notation:} Boldface letters refer to vectors (lower case) or matrices (upper case); $(\cdot)^T$, $(\cdot)^H$, $\|\cdot \|$, and $\|\cdot\|_{\max}$ denote transpose, complex conjugate transpose, Euclidean norm, and max-norm, respectively; $\textrm{diag}(\pmb{x})$ denotes a diagonal matrix whose main diagonal is $\pmb{x}$; $j=\sqrt{-1}$ is the imaginary unit; $\mathcal{R}(\cdot)$ and $\mathcal{I}(\cdot)$ return the real and imaginary part of their complex arguments, respectively; $\pmb{x}(k:n)$ refers to the vector with elements $x_{k},x_{k+1},\ldots,x_{n}$; the space of $x\times y$ complex matrices is denoted by $\mathbb{C}^{x\times y}$, $\mathbbm{1}$ denotes an all-ones column vector of appropriate dimension; $\textrm{sgn}(\cdot)$ denotes the sign function; $\mathcal{CN}(\mu,\sigma^2)$ denotes complex Gaussian distribution with mean $\mu$ and variance $\sigma^2$.

\section{Problem formulation}\label{sec1}

We consider a point-to-point MIMO system consisting of $N_T$ and $N_R$ transmit and receive antennas, respectively \cite{HAM}. Let $\pmb{H} \in \mathbbm{C}^{N_R\times N_T}$ denote the MIMO channel matrix, where entries correspond to the channels between the transmit and and receive antennas {\it i.e.}, $h_{i,j}$ is the channel coefficient between the $j$-th transmit antenna and the $i$-th receive antenna. Without loss of generality, we assume normalized Rayleigh block fading channels {\it i.e.,} $h_{i,j}\sim \mathcal{CN}(0,1)$. 

The MIMO system operates in the {\it beamforming mode} \cite{HAM} (single data flow) to maximize the received SNR and thus digital pre/post- processing vectors are adjusted accordingly. To reduce complexity and power consumption, both the transmitter and the receiver employ vector processing of $1$-bit resolution ({\it i.e.,} $1$-bit for the real part and $1$-bit for the imaginary part). Fig. \ref{fig_sys} schematically presents the system model.

Let $\pmb{f} \in \mathcal{S}^{N_T\times 1}$ and $\pmb{g} \in \mathcal{S}^{N_R\times 1}$ denote the (unnormalized) complex $1$-bit pre-coding and post-coding vectors, respectively, taking values in the quantized complex set $\mathcal{S}\triangleq \{\pm 1 \pm 1j\}$; the normalized pre/post-coding vectors are given by $\pmb{f}/\|\pmb{f}\|=\pmb{f}/\sqrt{2N_T}$ and $\pmb{g}/\|\pmb{g}\|=\pmb{g}/\sqrt{2N_R}$, respectively. We assume a perfect channel state information at both the transmitter and the receiver. The received SNR is given by

\begin{align}
\rho(\pmb{g},\pmb{f})=\frac{P|\pmb{g}^H\pmb{H}\pmb{f}|^2}{4 N_T N_R \sigma^2},\label{snr_ex}
\end{align}
where $\sigma^2$ is the variance of the additive white Gaussian noise, $P$ is the transmit power, while the extra terms in the denominator are due to the power normalization of the pre/post-coding vectors. Since the objective of the MIMO system is to maximize the received SNR, we introduce the following design problem 

\begin{subequations}
\begin{align}
&\max_{\pmb{f} \in \mathcal{S}^{N_T\times 1},\; \pmb{g} \in \mathcal{S}^{N_R \times 1}}\; |\pmb{g}^H \pmb{H}\pmb{f}|^2 \label{norm}  \\
&\;\;\;\;\;\;\;\;\;\;\;\;\;\;\; \Rightarrow \max_{\pmb{f} \in \mathcal{S}^{N_T\times 1},\; \pmb{g} \in \mathcal{S}^{N_R \times 1}}\; \pmb{f}^H(\pmb{H}^H\pmb{g}\pmb{g}^H \pmb{H})\pmb{f} \label{exp1}\\
&\;\;\;\;\;\;\;\;\;\;\;\;\;\;\; \Rightarrow \max_{\pmb{f} \in \mathcal{S}^{N_T\times 1},\; \pmb{g} \in \mathcal{S}^{N_R \times 1}}\; \pmb{g}^H(\pmb{H}\pmb{f}\pmb{f}^H\pmb{H}^H)\pmb{g}.  \label{exp2}
\end{align}
\end{subequations}
Due to the binary nature of the complex pre/post-coding vectors, the above combinatorial problem is NP-hard; the optimal solution requires ES over all the possible pre/post-coding vectors.

\subsection{Exhaustive search (ES)- Benchmark}\label{es1}

The ES scheme evaluates the SNR expression in \eqref{snr_ex} for all the possible pre/post-coding vectors and returns the solution with the maximum SNR. It is obvious that the ES requires $2^{2N_T}\times 2^{2N_R}=2^{2(N_T+N_R)}$ computations (complex valued vectors) and therefore its complexity becomes exponential with the number of transmit/receive antennas. A more intelligent ES scheme can take into account the computation symmetry embedded in \eqref{snr_ex}  {\it i.e.,} $\rho(\pmb{g},\pmb{f})=\rho(\pm \pmb{g},\pm \pmb{f})=\rho \big(-\mathcal{I}(\pmb{f})+\mathcal{R}(\pmb{f})j, +\mathcal{I}(\pmb{g})-\mathcal{R}(\pmb{g})j\big)=\rho \big(\mathcal{I}(\pmb{f})-\mathcal{R}(\pmb{f})j, -\mathcal{I}(\pmb{g})+\mathcal{R}(\pmb{g})j\big)$ corresponding to six equivalent computations and therefore the total number of ES computations decreases to $2^{2(N_T+N_R)}/6$; for both ES schemes, the implementation is prohibited for large-scale MIMO topologies.

\begin{algorithm}[t]
	\caption{QA-based $1$-bit pre/post-coding design.}\label{alg1}\vspace{1mm}
	\hspace*{\algorithmicindent}\textbf{Input:} $\pmb{H}$, $\pmb{g}_l^{(0)} \in \mathcal{S}^{N_R}$, $\pmb{f}_l^{(0)} \in \mathcal{S}^{N_T}$ with $l=1,\dots,L$, \hspace*{\algorithmicindent}relative tolerance $\delta$, $L$, $K$, $\rho_{\textrm{old}_l}= \rho(\pmb{g}_l^{(0)},\pmb{f}_l^{(0)})$, $k\leftarrow 0$.
	\begin{algorithmic}[1]
		\For{$l=1, 2,\ldots, L$} 
		\Repeat
		\State $k \leftarrow k + 1$
		\If{$k>1$}
		\State Let $\rho_{\textrm{old}_l}=\rho_{\textrm{new}_l}$
		\EndIf
		\State [D-WAVE] Solve $\pmb{b}_f^*\!\!=\!\!\arg \min_{\pmb{b}_f}\pmb{b}_f^T(-\pmb{V}_n)\pmb{b}_f$ for \hspace*{\algorithmicindent}\hspace*{\algorithmicindent} $\pmb{g}_l^{(k-1)}$.
		\State Convert $\pmb{b}_f^*$ to spin vector $\pmb{f}_l^{(k)}$. 
		\State [D-WAVE] Solve $\pmb{b}_g^*\!=\!\arg \min_{\pmb{b}_g}\pmb{b}_g^T(-\pmb{R}_n)\pmb{b}_g$ for \hspace*{\algorithmicindent}\hspace*{\algorithmicindent}$\pmb{f}_l^{(k)}$.
		\State Convert $\pmb{b}_g^*$ to spin vector $\pmb{g}_l^{(k)}$. 
		\State Let $\rho_{\textrm{new}_l}=\rho(\pmb{g}_l^{(k)},\pmb{f}_l^{(k)})$.
		\Until{$|\rho_{\textrm{new}_l}-\rho_{\textrm{old}_l}|/|\rho_{\textrm{old}_l}|<\delta$ or $k\geq L$}
		\State Obtain $(\pmb{g}_l,\pmb{f}_l)=(\pmb{g}_l^{(k)},\pmb{f}_l^{(k)})$.
		\EndFor
			\end{algorithmic}
		\hspace*{\algorithmicindent} \textbf{Output:} $(\pmb{g}^{\textrm{qa}},\pmb{f}^{\textrm{qa}})=\arg \max_{\pmb{g}_l,\pmb{f}_l} \rho(\pmb{g}_l,\pmb{f}_l)$.
\end{algorithm}

\section{QA-based $1$-bit pre/post-coding vector design}\label{sec2}

We propose a new algorithm that incorporates a QA solver in the design of the $1$-bit complex pre/post-coding vectors. By taking into account the binary combinatorial structure of the problem, we introduce an alternating optimization algorithm that solves a QUBO problem at each iteration. The QUBO formulations are solved through a state-of-the-art D-WAVE QA device \cite{WAVE}. More specifically, based on the equivalent expression in \eqref{exp1}, we firstly fix the vector $\pmb{g}$ (by using a random solution in the initial step) and we study the following combinatorial problem with respect to the pre-coding vector $\pmb{f}$ {\it i.e.,}
\begin{align}
\max_{\pmb{f} \in \mathcal{S}^{N_T\times 1}}\; \pmb{f}^H(\pmb{H}^H\pmb{g}\pmb{g}^H \pmb{H})\pmb{f}. \label{is1}
\end{align} 
Given that the involved matrices/vectors in \eqref{is1} are complex valued (which it is not compatible with current QA solvers), we transform the problem in \eqref{is1} to an equivalent real-valued form 

\begin{subequations}
\begin{align}
&\max_{\pmb{f}_r \in \mathcal{P}^{2N_T\times 1}} \pmb{f}_r^T(\pmb{J}^T \pmb{J})\pmb{f}_r,  \label{is11}\\
&\pmb{f}_r=\left[ \begin{array}{c} 
	\mathcal{R}(\pmb{f}) \\
	\mathcal{I}(\pmb{f}) \end{array} \right],\;\pmb{g}_r=\left[ \begin{array}{c} 
\mathcal{R}(\pmb{g}) \\
\mathcal{I}(\pmb{g}) \end{array} \right], \\
&\bar{\pmb{H}}=\left[ \begin{array}{cc} 
	\mathcal{R}(\pmb{H}) & \mathcal{I}(\pmb{H}) \\
	-\mathcal{I}(\pmb{H}) & \mathcal{R}(\pmb{H}) \end{array} \right],\; \pmb{q}=\pmb{g}_r^T\bar{\pmb{H}}, \\
&\pmb{J}=\left[ \begin{array}{cc} 
	\pmb{q}(1:N_R) & -\pmb{q}(N_R+1:2N_R) \\
	\pmb{q}(N_R+1:2N_R) & \pmb{q}(1:N_R) \end{array} \right],
\end{align}
\end{subequations}
where $\mathcal{P}\triangleq\{+1,-1\}$ denotes the spin set. Then, we convert the pre-coding vector $\pmb{f}_r$ (containing spin variables $\{+1,-1\}$) to the binary vector $\pmb{b}_f$ (with entries in the set $\{0,1\}$) by using the transformation \cite{JAM1} $\pmb{b}_f=\frac{1}{2}(\pmb{f}_r+\mathbbm{1})$ and thus the problem in \eqref{is11} is converted to a standard QUBO formulation
\begin{align}
&\max_{\pmb{f}_r \in \mathcal{P}^{2N_T\times 1}} \pmb{f}_r^T(\pmb{J}^T \pmb{J})\pmb{f}_r= \max_{\pmb{f}_r \in \mathcal{P}^{2N_T\times 1}} \pmb{f}_r^T\pmb{V}\pmb{f}_r \\
&=\max_{\pmb{b}_f}\;4 \pmb{b}_f^T \pmb{V}\pmb{b}_f-2\pmb{b}_f^T\pmb{V}\mathbbm{1}-2\mathbbm{1}^T\pmb{V}\pmb{b}_f \nonumber \\
&=\max_{\pmb{b}_f}\; \pmb{b}_f^T \bigg( 4\pmb{V}-4\textrm{diag}(\pmb{V}\mathbbm{1}) \bigg)\pmb{b}_f \nonumber \\
&=\max_{\pmb{b}_f}\; \pmb{b}_f^T \pmb{V}_0 \pmb{b}_f.
\end{align}
The last mathematical step transforms the above (symmetric) quadratic matrix $\pmb{V}_0$ to a form that is compatible with the D-WAVE QA solver \cite{JAM2}; we normalize the matrix such as all its entries take values in the range $[-1,+1]$. The calibrated matrix is written by $\pmb{V}_n=\frac{\pmb{V}_0}{\| \pmb{V}_0\|_{\max}}$ and thus the D-WAVE compatible QUBO problem is written as
\begin{align}
\min_{\pmb{b}_f}\;\pmb{b}_f^T (-\pmb{V}_n) \pmb{b}_f. \label{qubo1}
\end{align}
The returned binary vector $\pmb{b}_f$ is converted back to the spin vector $\pmb{f}$. By using the principles of alternating optimization, we then fix $\pmb{f}$ (by using the previous solution), and we solve the equivalent problem in \eqref{exp2} {\it i.e.,}
\begin{align}
\max_{\pmb{g} \in \mathcal{S}^{N_R \times 1}}\; \pmb{g}^H(\pmb{H}\pmb{f}\pmb{f}^H\pmb{H}^H)\pmb{g}, 
\end{align}
which is transformed to an equivalent real-valued spin form 
\begin{subequations}
\begin{align}
&\max_{\pmb{g}_r \in \mathcal{P}^{2N_T\times 1}} \pmb{g}_r^T(\pmb{Z}^T \pmb{Z})\pmb{g}_r  \label{is22}\\
&\hat{\pmb{H}}=\left[ \begin{array}{cc} 
	\mathcal{R}(\pmb{H}) & -\mathcal{I}(\pmb{H}) \\
	\mathcal{I}(\pmb{H}) & \mathcal{R}(\pmb{H}) \end{array} \right],\; \pmb{z}=\hat{\pmb{H}}\pmb{f}_r, \\
&\pmb{Z}=\left[ \begin{array}{cc} 
	\pmb{z}(1:N_T) & \pmb{z}(N_T+1:2N_T) \\
	-\pmb{z}(N_T+1:2N_T) & \pmb{z}(1:N_T) \end{array} \right]^T.
\end{align}
\end{subequations}
The above spin-variable formulation is converted to a QUBO formulation by using similar analytical steps like before 
\begin{align}
&\max_{\pmb{g}_r \in \mathcal{P}^{2N_T\times 1}} \pmb{g}_r^T(\pmb{Z}^H \pmb{Z})\pmb{g}_r\;=\max_{\pmb{g}_r \in \mathcal{P}^{2N_T\times 1}} \pmb{g}_r^T \pmb{R}\pmb{g}_r\\
&=\min_{\pmb{b}_g}\; \pmb{b}_g^T(-\pmb{R}_n)\pmb{b}_g,
\end{align}
where $\pmb{b}_g$ is the binary representation of the vector $\pmb{g}$, while the normalized matrix $\pmb{R}_n$ is defined similar to $\pmb{V}_n$. 

The above alternating optimization process is repeated until convergence or a maximum number of iterations ($K$ iterations) is achieved. Due to the discrete/binary nature of the problem, the final solution is sensitive to the initial conditions ({\it i.e.,} initial vector $\pmb{g}^{(0)})$ which could result in convergence to a local maximum (SNR objective function). To overcome this limitation and facilitate the iterative algorithm to escape local maxima and get closer to the optimal solution, we introduce $L$ independent execution of the algorithm by using different initial conditions ($\pmb{g}_l^{(0)}$ with $l=1,\ldots,L$); from the $L$ returned solutions, we keep the one corresponding to the maximum SNR objective function. The pseudocode of the QA-based iterative algorithm is given in Algorithm \ref{alg1}.  

The time complexity of the Algorithm \ref{alg1} mainly refers to the number of anneals, which is a system parameter that is tuned empirically. Obviously, as the size of the problem increases, the number of anneals should also increase (in a polynomial way) in order to successfully determine the optimal solution. Specifically, the time complexity of the proposed QA scheme can be written as $K\times L\times (2\times N_a)\times T_a$ (time-units) where $N_a$ is the number of anneals, $T_a$ is the anneal time, while the factor $2$ is due to the fact that we solve two QUBO problems at each basic algorithmic iteration. It is obvious that the time complexity of the proposed scheme is polynomial with respect the main system parameters and thus much lower than the exponential time complexity of the classical ES scheme.

\begin{figure}
	\centering
	\includegraphics[width=0.9\linewidth]{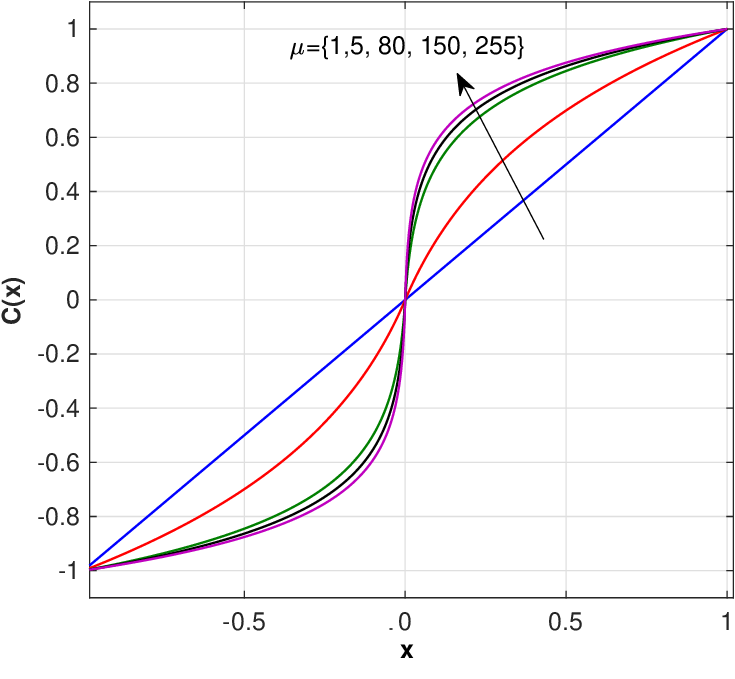}
	\caption{$\mu$-law compander characteristic function; $\mu=255$ is considered in our experimental studies (commercial PCM).}\label{fig1}
\end{figure}

\begin{figure}
	\centering
	\includegraphics[width=0.8\linewidth]{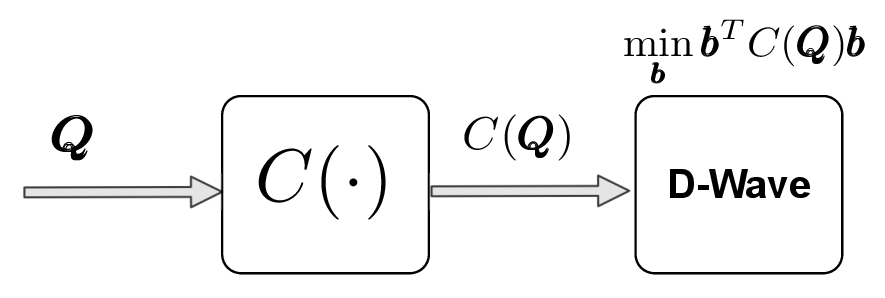}
	\caption{QUBO pre-processing technique to combat Hamiltonian/ICE noise.}\label{fig2}
\end{figure}   

\begin{figure*}[h]
	\centering
	\includegraphics[width=0.7\linewidth]{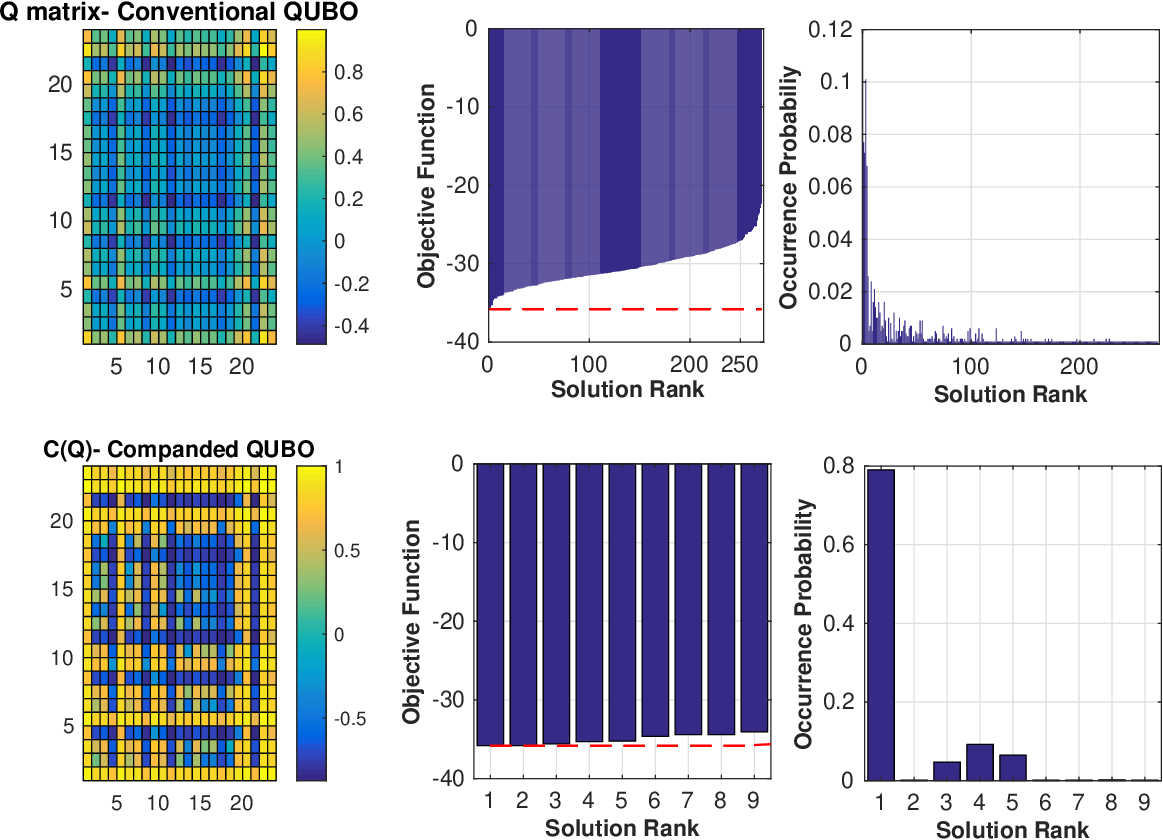}
	\caption{{\bf [top]} D-WAVE performance for conventional QUBO (without pre-processing), {\bf [bottom]} D-WAVE performance for companded QUBO (with pre-processing); QUBO problem of dimension $24$, ES benchmark (dashed line).}\label{fig3}
\end{figure*}

\section{Pre-processing- companded QUBO}\label{sec3}

In this section, we study a new pre-processing QA technique that converts the quadratic QUBO matrix in a form that is more robust to Hamiltonian noise \cite{ZAB}. Due to the interaction of the D-WAVE QA device with the external environment as well as due to hardware limitations, the QA process suffers from noise effects that decrease the quality of the solution. These noise effects are collectively represented by the term integrated control errors (ICE) which captures all the sources of solution infidelity ({\it e.g.,} cross-talk with adjacent qubits, temperature fluctuations, flux noise, quantization/precision, bandwidth of the analogue circuits {\it etc.}). An informative overview of the ICE sources is given in \cite[Ch. 3]{ICE}. Due to the ICE effects, the QA device solves a slightly altered QUBO problem given by 
\begin{align}
\underbrace{\min_{\pmb{b}}\;\pmb{b}^T\pmb{Q} \pmb{b}}_{\text{original QUBO}}\;\rightarrow\; \underbrace{\min_{\pmb{b}}\;\pmb{b}^T\big(\pmb{Q}+\pmb{E}\big) \pmb{b}}_{\textrm{noisy/ICE QUBO}}, \label{QUBOmod}
\end{align}   
where $\pmb{Q}$ is the original symmetric quadratic QUBO matrix, while the random matrix $\pmb{E}$ represents the Hamiltonian/ICE noise in the bias and couplers coefficients (the elements of the matrix $\pmb{E}$ are random variables {\it e.g.,} coloured Gaussian distribution \cite{ICE}). Therefore, it is obvious that the QA device solves a modified/noisy problem whose ground state may be different that the one of the original QUBO problem. The associated literature mainly focuses on the characterization of the noise by using statistical signal processing tools \cite{ZAB} or the investigation of heuristic post-processing techniques that improve the quality of the returned solution \cite{DOR}. In practice, to combat the noise/ICE effects and overcome potential local minima, the same QUBO instance is solved multiple times (each execution is called {\it anneal}) in a row and the best solution among all the anneals is adopted\footnote{It is worth noting that this concept is very similar to retransmission techniques ({\it e.g.,} repetition code) in wireless communications systems, where the same information signal is transmitted multiple times through independent noisy channels to achieve diversity and increase reliability \cite[Ch. 3]{TSE}.}.  

From the modified QUBO problem in \eqref{QUBOmod}, it is obvious that the elements of the original matrix $\pmb{Q}$ with small values/magnitudes ($\approx 0$) are more sensitive to noise effects. Our proposed pre-processing QUBO technique converts the $\pmb{Q}$ matrix elements with small magnitude to higher values without affecting the qualitative interconnection between the qubits ({\it i.e.,} the sign and the order in the strength of the bias/couplers coefficients remain the same). Specifically, inspired by the non-uniform quantization process in PCM systems which boosts the lower amplitudes and reduces the dynamic range of the input speech signal \cite[Ch. 7]{PRO}, we introduce a nonlinear compression block (named {\it compander} by using the terminology from PCM systems) to pre-process $\pmb{Q}$ elements before the D-WAVE QA operation. Without loss of generality, we consider the $\mu$-low (logarithmic) compander which is used in PCM systems in the United States and Canada, defined as follows \cite[Ch. 7]{PRO}
\begin{align}
C(x)=\frac{\log(1+\mu|x|)}{\log(1+\mu)}{\textrm{sgn}(x)},\;\;\;\textrm{with}\;-1 \leq x\leq 1.
\end{align}   
The positive parameter $\mu$ controls the amount of compression. Fig. \ref{fig1} plots the characteristic function of the $\mu$-law for various values of $\mu$; it can be seen that as the $\mu$ parameter increases, the low magnitude elements are mapped to much higher values. In our experimental studies (without loss of generality), we consider $\mu=255$ which is the value that is used in commercial PCM systems. After QUBO pre-processing, the D-WAVE QA device solves the following noise-robust QUBO problem     
\begin{align}
	\min_{\pmb{b}}\;\pmb{b}^T\bigg(C(\pmb{Q})+\pmb{E}\bigg) \pmb{b} \label{QUBOmod2}.
\end{align}

Fig. \ref{fig2} schematically presents the proposed QA pre-processing technique to combat ICE effects; it is worth noting, that the proposed approach is generic and be used with other physics/quantum-inspired heuristic solvers ({\it e.g.} coherent Ising machines).

\section{Experimental evaluation}\label{sec4}

Experimental results in a state-of-the-art D-WAVE QA device are carried-out in order to validate the efficiency of the proposed techniques. 

\subsection{D-WAVE setting}

For our D-WAVE QA experiments, we use the D-WAVE Leap interface with the {\it Advantage\_system1.1} quantum processing unit \cite{WAVE}. For each QUBO instance, we use $1,000$ anneals with $1\mu$sec anneal time and a ferromagnetic coupling parameter equal to $\bar{J}_F=3$. For the minor embedding, we adopt the heuristic algorithm {\it minorminer} which is included in the Ocean software development kit by default \cite{KAS,WAVE}; in this case, a majority vote is applied to broken chains.

\subsection{Companded QUBO}

In Fig. \ref{fig3}, we study the efficiency of the proposed QUBO pre-processing technique. As an indicative example, we consider a QUBO problem (see eq. \eqref{QUBOmod}) of dimension $24$ where the elements of the QUBO symmetric matrix $\pmb{Q}$ are random variables of normal Gaussian distribution with variance one (a normalization in the range $[-1,+1]$ is also applied $\pmb{Q}/\|\pmb{Q}\|_{\max}$). In Fig. \ref{fig3}[top], we plot the performance of the conventional QUBO (without pre-processing); specifically, we plot the original QUBO matrix as an array of coloured cells, the returned solutions over $1,000$ anneals which are ordered in descending order of their objective function values as well as the associated occurrence probabilities. As it can be seen, the D-WAVE solver returns almost $270$ distinct solutions and the optimal solution (equivalent to ES scheme) occurs with a probability $\approx 0.08$. We also note that the considered QUBO problem requires $24$ logical variables corresponding to (average) $80$ physical qubits for the minor embedding.

In Fig. \ref{fig3}[bottom], we show the performance of the proposed companded QUBO. Firstly, we plot the companded QUBO matrix $C(\pmb{Q})$; it can be seen the low magnitude elements of the original matrix $\pmb{Q}$ are converted to higher values. As for the efficiency of the pre-processing process, it can be seen that the D-WAVE device returns only $9$ (high quality) solutions while the optimal solution (equivalent to ES) occurs with a probability $\approx 0.8$. The obtained results show that the proposed pre-processing technique preserves the quadratic QUBO matrix from the Hamiltonian noise and increases the efficiency of the D-WAVE solver. The fact that the probability of the optimal solution increases by a factor $10$ could decrease the required number of the total anneals, making the D-WAVE QA solver more attractive for real-time (delay-sensitive) applications.  

\begin{figure}
	\centering
	\includegraphics[width=\linewidth]{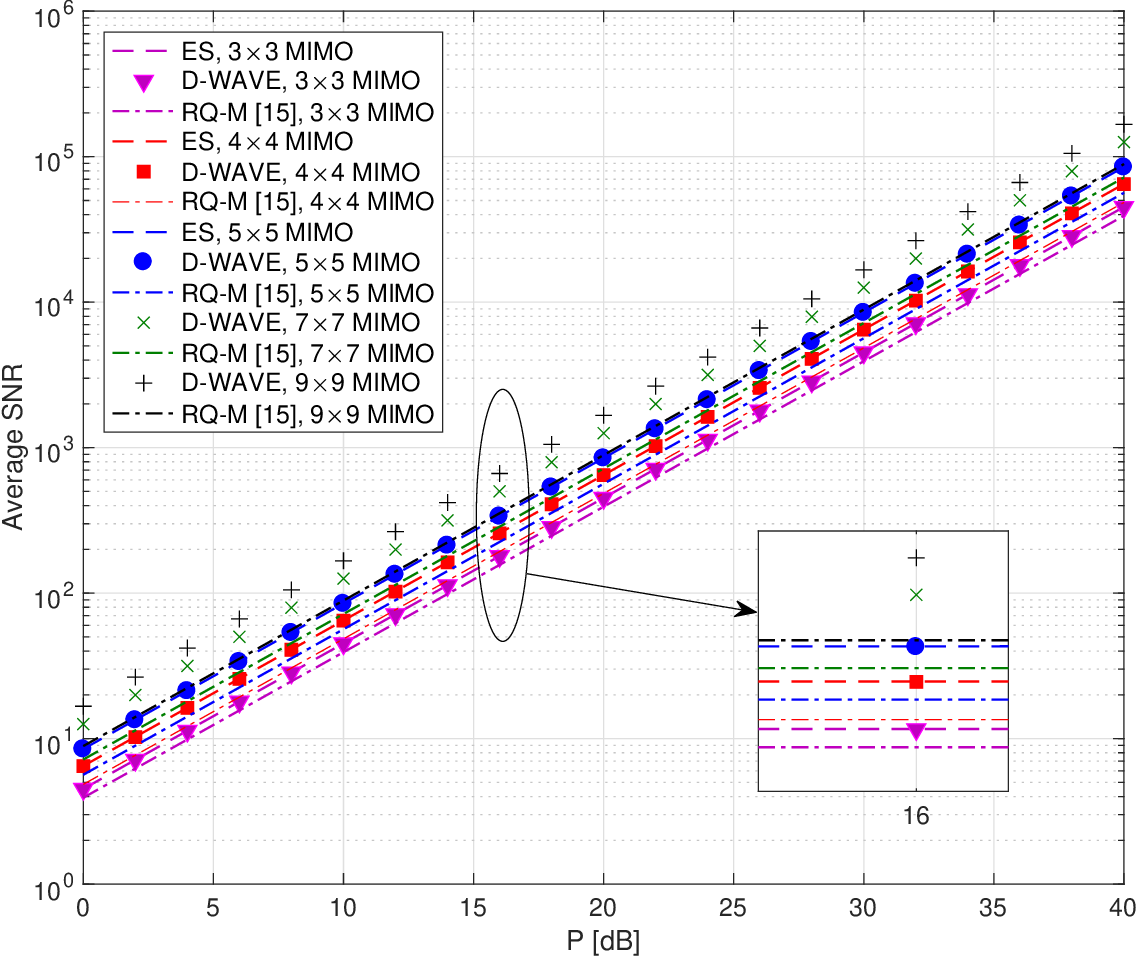}
	\caption{[D-WAVE] Average SNR performance versus transmit power $P$ for different MIMO configurations; Algorithm \ref{alg1}/D-WAVE results (markers), ES benchmark (dashed line), RQ-M algorithm \cite{KRI3} (dashdotted line).}\label{fig4}
\end{figure}

\begin{figure}
	\centering
	\includegraphics[width=0.8\linewidth]{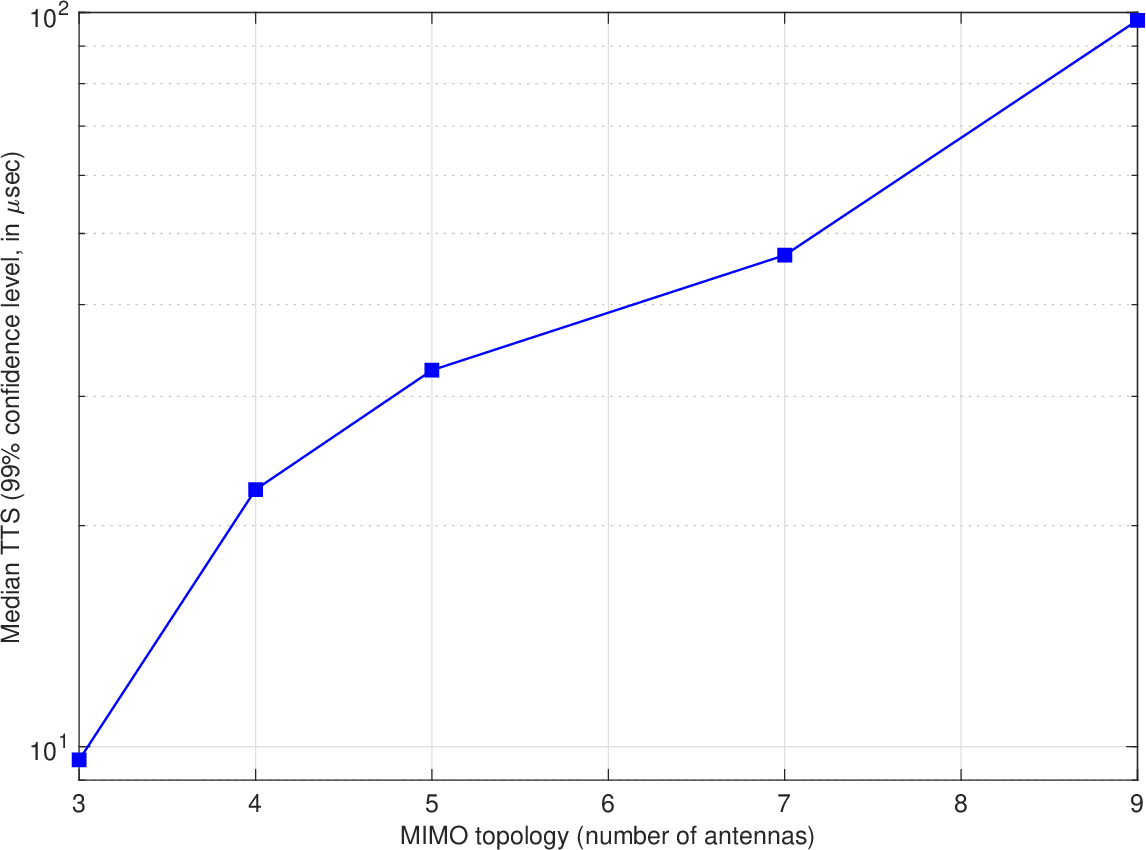}
	\caption{Median TTS ($99$\% confidence level) for a single D-WAVE iteration, versus the size of the MIMO topology with $N_T=N_R$.).}\label{fig55}
\end{figure}

\subsubsection{Minimum spectral gap}

To further demonstrate the efficiency of the proposed pre-processing technique, we study the minimum spectral (eigenvalue) gap of the QA adiabatic evolution for some indicative problem instances. Specifically, the quantum Hamiltonian that drives the D-WAVE QA process is given by
\begin{align}
H(s)&=-\frac{A(s)}{2}\bigg(\sum_i \hat{\sigma}_x^{(i)} \bigg)+\frac{B(s)}{2}\bigg(\sum_{i}h_i\hat{\sigma}_z^{(i)} \nonumber \\
&\;\;\;\;\;\;\;\;\;\;\;\;\;\;\;\;\;\;\;\;\;\;\;\;\;\;\;\;\;\;\;\;\;+\sum_{i<j}J_{i,j}\hat{\sigma}_z^{(i)}\hat{\sigma}_z^{(j)} \bigg),
\end{align}
where $\hat{\sigma}_{x,z}^{(i)}$ are the Pauli matrices operating on a qubit $q_i$, the coefficients $h_i$ and $J_{i,j}$ are the linear biases and quadratic couplers of the equivalent Ising problem, respectively, $s \in [0\; 1]$ denotes the anneal fraction, $A(s)$ and $B(s)$ represent the anneal schedules (adiabatic path). The values of $A(s)$ and $B(s)$ for each value of the anneal fraction are available for the latest D-WAVE quantum processors in \cite{ADI}; this allows us to simulate the Hamiltonian evolution (with an accuracy $30\%$ \cite{ADI}) of the quantum system and characterize the minimum spectral gap that governs its efficiency \cite{MCG}. We recall that the the minimum spectral gap is defined as the minimum of the energy difference between the first excited energy and the ground energy as a function of the time {\it i.e.,} 
\begin{align}
g=\min_{s \in [0,\;1]} \lambda_1(s)-\lambda_0(s),
\end{align}
where $\lambda_i(s)$ denotes the $i$-th eigenvalue of the Hamiltonian $H(s)$ at the time fraction $s$ with $\lambda_0 \leq \lambda_1 \leq \ldots \leq \lambda_{2^n}$, where $n$ is the number of qubits in the quantum system.

\begin{figure}
	\centering
	\includegraphics[width=\linewidth]{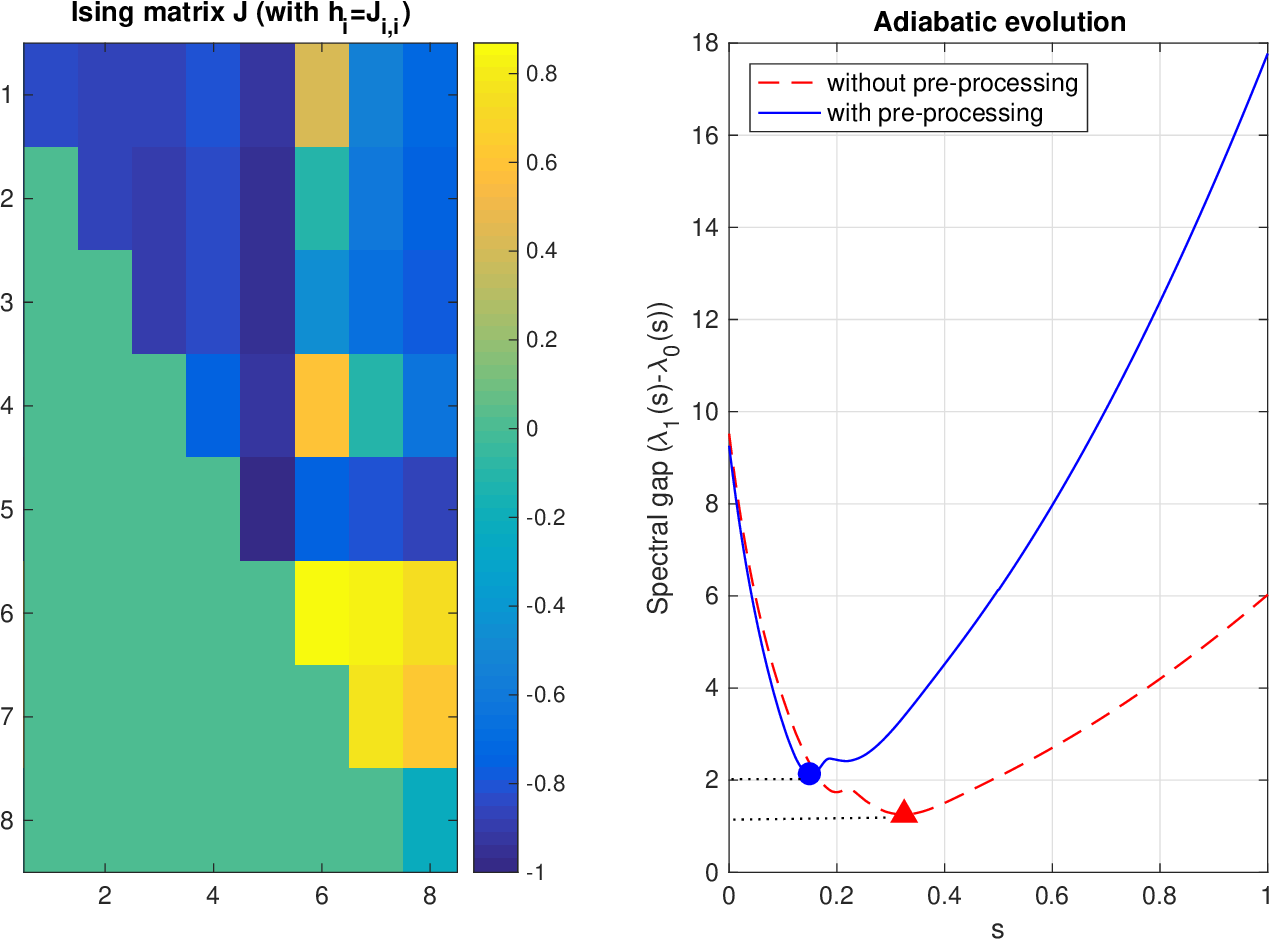}
	\vspace{-0.3cm}
	\caption{{\bf [left]} Ising matrix, {\bf{[right]}} Spectral minimum gap versus the anneal fraction. }\label{fig44}
\end{figure}

In Fig. \ref{fig44}, we plot the spectral gap  versus the anneal fraction, for an Ising problem with $n=8$ qubits where the parameters $h_i, J_{i,j}$ are distributed according to a Gaussian normal distribution with variance one (normalized to the range $[-1,\;+1]$). It can be seen that the proposed companded pre-processing technique ensures a higher minimum spectral gap ({\it i.e.,} $g=2,11$ against $g=1,25$ without pre-processing for the considered example), which validates its robustness against the Hamiltonian noise. It is also worth noting that a higher minimum spectral gap can minimize the overall computation anneal time, since the appropriate anneal duration is proportional to $T_{a}\sim 1/g^2$. 

In Table \ref{tab1}, we study the efficiency of the pre-processing technique in terms of average minimum spectral gap $\mathbbm{E}(g)$ over $10,000$ independent Rayleigh fading channel realizations. It can be seen that the proposed pre-processing technique achieves a higher average minimum spectral gap for all scenarios considered while it ensures a high efficiency ({\it i.e.,} pre-processing efficiency denotes the probability that the minimum spectral gap (with pre-processing) is higher than this one without pre-processing).

\begin{table}[h]
\caption{Average minimum spectral gap $\mathbbm{E}(g)$.}\label{tab1}
\vspace{-0.1cm}
\centering
\begin{tabular}{|c||c||c||c|}
\hline
 & $n=5$ & $n=8$ & $n=10$ \\
\hline 
{\bf Without pre-processing} & $0,8906$ & $0,8221$ & $0,7750$ \\
\hline
{\bf With pre-processing} & $1,2377$ & $1,2469$ & $1,2612$ \\
\hline
{\bf Pre-processing efficiency ($\%$)} & $77,56$ & $75,67$ & $76,21$ \\
\hline
\end{tabular}
\end{table}


\subsection{MIMO $1$-bit pre/post-coding design}

We study the performance of the proposed QA-based $1$-bit pre/post-coding MIMO design. We consider a symmetric MIMO setup with $N_T=N_R$, $\sigma^2=1$, $K=8$, $L=8$, $\delta=0.01$. In Fig. \ref{fig4}, we plot the average SNR performance of the proposed algorithm (Algorithm \ref{alg1}) versus the transmit power $P$; we consider $1,000$ channel realizations while the ES solution and the RQ-M algorithm \cite{KRI3} (classical heuristic which is based on the Rayleigh quotient) are used as performance benchmarks. It can be seen, that the proposed QA-based iterative algorithm achieves the optimal ES performance for all cases. The RQ-M scheme provides a suboptimal performance mainly due to the discretization of the complex-valued solution (quantization noise) \cite{KRI3}. It is worth noting that as the number of antennas increases, ES can not be applied since the total number of computations becomes practically prohibited {\it e.g.,} $6,87\times 10^{10}$  computations per channel realization for the scenario with $N_T=N_R=9$. The MIMO topologies considered have high practical interest for the upcoming communication systems and are sufficient to demonstrate the efficiency of the proposed QA-based framework.

In Fig. \ref{fig55}, we plot the median time-to-solution (TTS) versus the size of the MIMO topology for a single D-WAVE iteration (find $\pmb{f}$ or $\pmb{g}$). The TTS is defined as the time needed for the D-WAVE solver to find the ground state of the problem with $99\%$ success probability {\it i.e.}, $\textrm{TTS}=T_{\text{run}} \log(0.01)/\log(1-p)$ where $t_{\text{run}}$ is the total annealing time ($1\mu sec$ for $1,000$ anneals) and $p$ is the success probability. We observe that as the MIMO size increases, a higher TTS is required to find the optimal pre/post coding vectors; TTS increases by a factor $10$, as we move from a $3\times 3$ MIMO to a $9\times 9$ MIMO topology. 

\begin{figure}
	\centering
	\includegraphics[width=\linewidth]{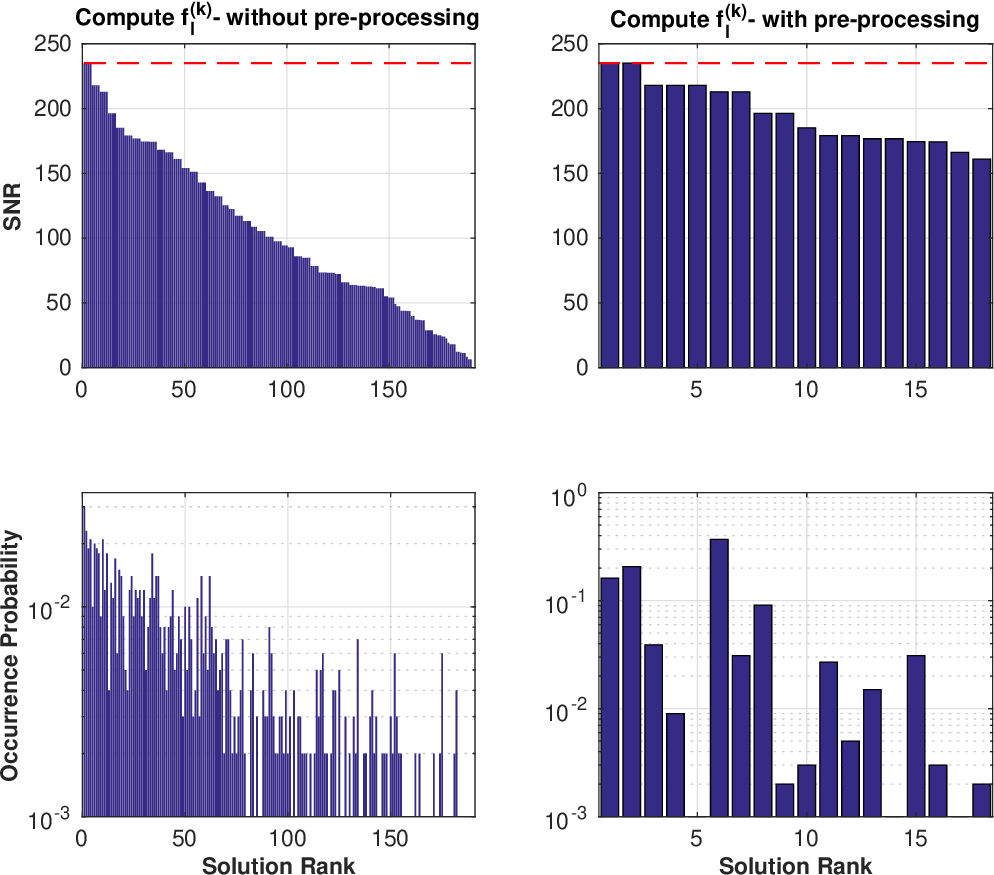}
	\caption{D-WAVE indicative results for a single channel and single iteration to compute $\pmb{f}_l^{(k)}$; SNR performance of the returned solutions (in descending order) and the associated occurrence probabilities;  ES benchmark (dashed line), $N_R=N_T=4$, $P=18$ dB. {\bf [left]} conventional QUBO (without pre-processing), {\bf [right]} proposed companded QUBO (with pre-processing).}\label{fig5}
\end{figure}

\begin{figure}
	\centering
	\includegraphics[width=\linewidth]{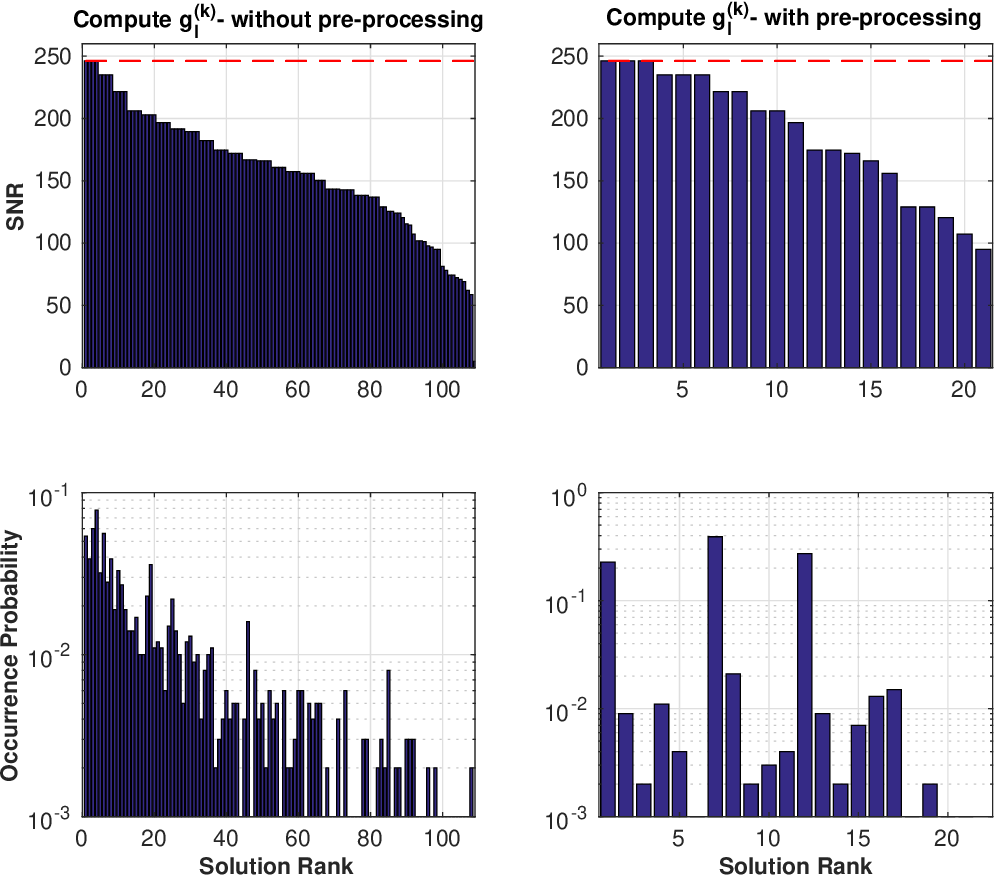}
	\caption{D-WAVE indicative results for a single channel and single iteration to compute $\pmb{g}_l^{(k)}$; SNR performance of the returned solutions (in descending order) and the associated occurrence probabilities;  ES benchmark (dashed line), $N_R=N_T=4$, $P=18$ dB. {\bf [left]} conventional QUBO (without pre-processing), {\bf [right]} proposed companded QUBO (with pre-processing).}\label{fig6}
\end{figure}

In Figures \ref{fig5}, \ref{fig6}, we focus on a single iteration of the proposed algorithm (Algorithm \ref{alg1}) for a single channel realization for a setup with $N_T=N_R=4$, $P=18$ dB. Specifically, Fig. \ref{fig5}[left] deals with the design of the pre-coding vector $\pmb{f}_l^{(k)}$ (given a post-coding vector $\pmb{g}_l^{(k-1)}$) without QUBO pre-processing. Firstly, we observe that the D-WAVE returns almost $\approx 190$ solutions while the optimal solution (with the maximum SNR) occurs with a total probability $0.09$; it is worth noting that the first four (in the order) solutions are equivalent in terms of SNR performance, due to the computation symmetry of the problem (see Section \ref{es1}). In Fig. \ref{fig5}[right], we adopt the proposed QUBO pre-processing technique and we show the performance of the D-WAVE in terms of SNR for the returned solutions as well as the associated occurrence probabilities. The results validate our previous main observations {\it i.e.,} the companded pre-processing QUBO approach improves the quality of the returned solutions (only $18$ high quality solutions are returned), while the optimal solution occurs with a probability $\approx 0.37$. 

Fig. \ref{fig6} deals with the next step of the iterative algorithm which refers to the design of the post-coding vector $\pmb{g}_l^{(k)}$ given the post-coding vector from the previous step. The results are inline with our previous remarks; the D-WAVE solver achieves the optimal ES performance while the companded QUBO pre-processing improves the quality of the returned solutions and slightly increases the occurrence probability of the optimal solutions. It is also worth noting that the achieved maximum SNR increases at the end of the second step in comparison to the SNR output of the previous algorithmic step.

\section{Conclusion}

In this paper, we studied a new low-complexity digital MIMO architecture with $1$-bit (processing) resolution per complex dimension. The design of pre/post-coding that maximizes the received SNR relies to an NP-hard combinatorial problem with exponential complexity for the optimal ES scheme. An iterative D-WAVE QA-based algorithm that solves appropriate real-valued QUBO instances at each iteration is proposed; the iterative scheme gradually achieves near-optimal (similar to ES) performance while ensuring polynomial complexity. To further boost D-WAVE efficiency,  a pre-processing non-linear companding technique that transforms the quadratic matrix to a form that is more robust to Hamiltonian/ICE noise is also investigated. Experimental results show that the new pre-processing technique improves the quality of the D-WAVE solutions and could decrease the required number of the total anneals.

\begin{IEEEbiography}[{\includegraphics[width=1in,height=1.25in,clip,keepaspectratio]{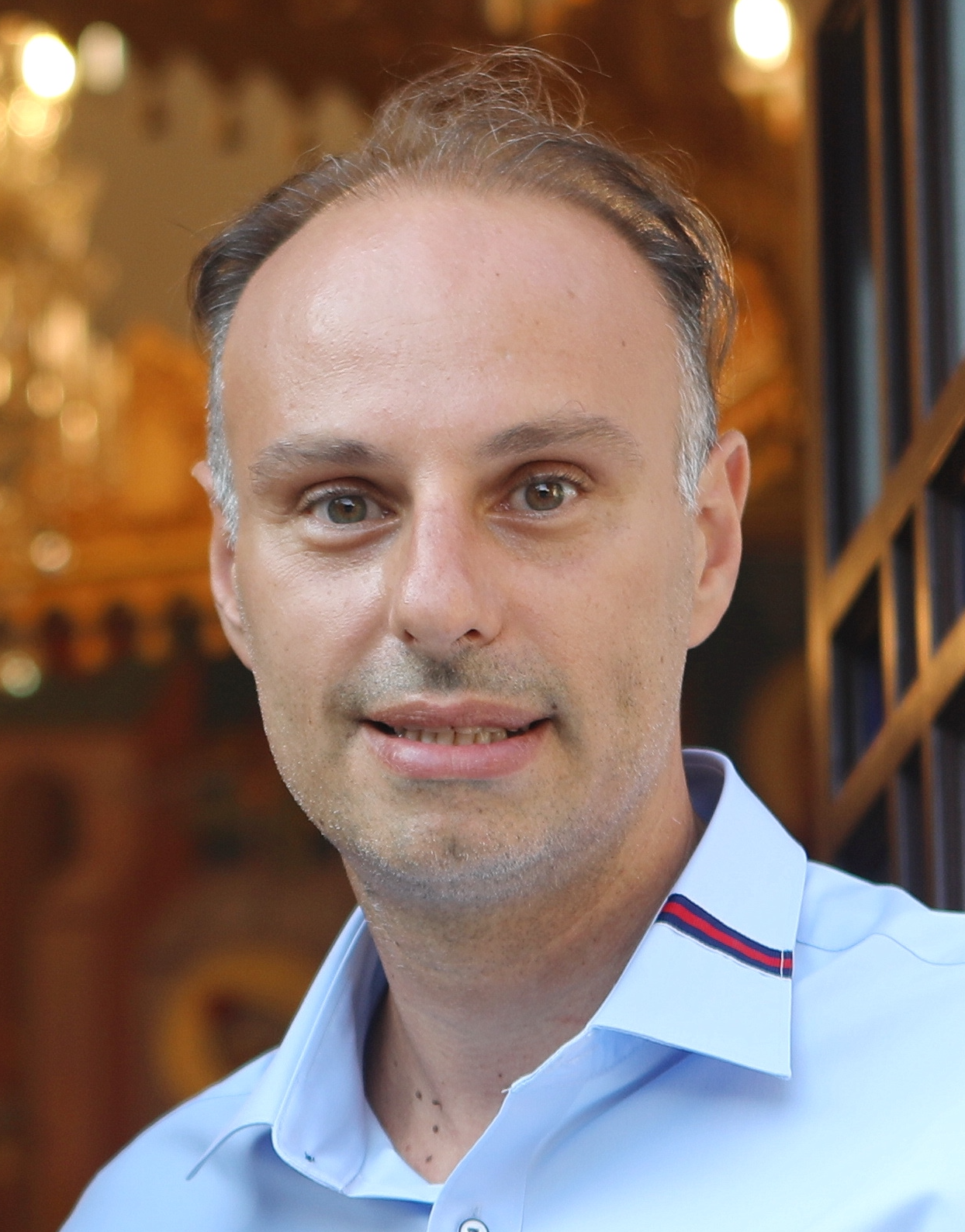}}]{Ioannis Krikidis} (F’19) received the diploma in Computer Engineering from the Computer Engineering and Informatics Department (CEID) of the University of Patras, Greece, in 2000, and the M.Sc and Ph.D degrees from \'Ecole Nationale Sup\'erieure des T\'el\'ecommunications (ENST), Paris, France, in 2001 and 2005, respectively, all in Electrical Engineering. From 2006 to 2007 he worked, as a Post-Doctoral researcher, with ENST, Paris, France, and from 2007 to 2010 he was a Research Fellow in the School of Engineering and Electronics at the University of Edinburgh, Edinburgh, UK. 

He is currently a Professor at the Department of Electrical and Computer Engineering, University of Cyprus, Nicosia, Cyprus. His current research interests include wireless communications, quantum computing, 6G communication systems, wireless powered communications, and intelligent reflecting surfaces. Dr. Krikidis serves as an Associate Editor for IEEE Transactions on Wireless Communications, and Senior Editor for IEEE Wireless Communications Letters. He was the recipient of the Young Researcher Award from the Research Promotion Foundation, Cyprus, in 2013, and the recipient of the IEEEComSoc Best Young Professional Award in Academia, 2016, and IEEE Signal Processing Letters best paper award 2019. He has been recognized by the Web of Science as a Highly Cited Researcher for 2017-2021. He has received the prestigious ERC Consolidator Grant for his work on wireless powered communications. 
\end{IEEEbiography}

\EOD


\begin{thebibliography}{00}


\bibitem{ZHE} Z. Wang, J. Zhang, H. Du, D. Niyato, S. Cui, B. Ai, M. Debbah, K. B. Letaief, and H. V. Poor, ``A tutorial on extremely large-scale MIMO for 6G: fundamentals, signal processing, and applications,'' {\it IEEE Commun. Surv. Tut.}, (to appear), 2024. 

\bibitem{ALI} A. Li, C. Masouros, A. L. Swindlehurst, and W. Yu, ``1-bit massive MIMO transmission: Embracing interference with symbol-level precoding,'' {\it IEEE Comm. Mag.}, vol. 59, pp. 121--127, May 2021. 

\bibitem{SIL} J. M. B. da Silva, A. Sabharwal, G. Fodor, and C. Fischione, ``1-bit phase shifters for large-antenna full-duplex mmWave communications,'' {\it IEEE Trans. Wireless Commun.}, vol. 19, pp.  6916--6931, Oct. 2020.


\bibitem{LOZ} A. Lozano, ``$1$-bit MIMO for Terahertz channels,'' in {\it Proc. Int. ITG Work. Smart Antennas}, French Riviera, France, Nov. 2021. 



\bibitem{MCG} C. C. McGeoch, {\it Adiabatic quantum computation and quantum annealing}, Morgan \& Claypool Publ., 2014.

\bibitem{KAS} S. Kasi, P. A. Warburton, J. Kaewell, and K. Jamieson, ``A cost and power feasibility analysis of quantum annealing for nextG cellular
wireless networks,'' {\it IEEE Trans. Quantum Eng.}, vol. 4, pp. 1--17, Oct. 2023.

\bibitem{WAVE} D-WAVE, {\it Getting started with D-WAVE solvers}, \url{https://docs.ocean.dwavesys.com/en/stable/getting_started.html}, 2021.


\bibitem{KIM} M. Kim, S. Kasi, P. A. Lott, D. Venturelli, J. Kaewell, and K. Jamieson, ``Heuristic quantum optimization for 6G wireless communications,'' {\it IEEE Networks}, vol. 35, pp. 8--15, July 2021. 


\bibitem{JAM1} M. Kim, D. Venturelli, and K. Jamieson, ``Leveraging quantum annealing for large MIMO processing in centralized radio access networks,'' in
{\it Proc. ACM Special Interest Group Data Commun.}, Beijing, China, Aug. 2019, pp. 241--255. 

\bibitem{JAM2} A. K. Singh, D. Venturelli, and K. Jamieson, ``Perturbation-based
formulation of maximum likelihood MIMO detection for coherent ising
machines,'' in {\it Proc. IEEE Global Commun. Conf.}, Rio de Janeiro, Brasil,
Dec. 2022, pp. 2523--2528.

\bibitem{KASI} S. Kasi, J. Kaewell, and K. Jamieson, ``A quantum annealer-enabled decoder and hardware topology for nextG wireless polar codes,'' {\it IEEE Trans. Wireless Commun.}, (to appear), 2023. 


\bibitem{ROSI} C. Ross, G. Gradoni, Q. J. Lim, and Z. Peng, ``Engineering reflective metasurfaces with Ising Hamiltonian and quantum annealing,'' {\it IEEE Trans. Antennas Propag.}, vol. 70, pp. 2841--2854, April 2022. 


\bibitem{KRI} I. Krikidis, C. Psomas, A. K. Singh, and K. Jamieson, ``Optimizing Configuration Selection in Reconfigurable-Antenna MIMO Systems: Physics-Inspired Heuristic Solvers,'' {\it IEEE Trans. Commun.}, (submitted), Jan. 2024. 

\bibitem{DIN} T. Q. Dinh, S. H. Dau, E. Lagunas, and S. Chatzinotas, ``Efficient Hamiltonian reduction for quantum annealing on Satcom beam placement problem,'' in {\it Proc. IEEE Int. Conf. Commun.}, Rome, Italy, May 2023, pp. 2668--2673. 

\bibitem{KRI3} I. Krikidis, ``MIMO with analogue 1-bit phase shifters: A quantum annealing perspective'' {\it IEEE Wireless Commun. Lett.}(to appear), 2024. 


\bibitem{ICE} D-WAVE, {\it Technical description of the D-WAVE quantum processing unit}, User manual, Feb. 2023. 


\bibitem{PRO} J. G. Proakis and M. Salehi, {\it Fundamentals of Communication systems}, Pearson Prentice Hall, 2005. 


\bibitem{HAM} J. R. Hampton, {\it Introduction to MIMO communications}, Cambridge, 2014.


\bibitem{ZAB} T. Zaborniak and R. de Sousa, ``Benchmarking Hamiltonian noise in the D-WAVE quantum annealer,'' {\it IEEE Trans. Quantum Eng.}, vol. 2, No. 3100206, Jan. 2021. 


\bibitem{DOR} J. E. Dorband, ``Improving the accuracy of an adiabatic quantum computer'', \url{https://arxiv.org/abs/1705.01942}, 2017. 


\bibitem{TSE} D. Tse and P. Viswanath, {\it Fundamentals of wireless communication}, Cambridge, 2005. 

\bibitem{ADI} \url{https://docs.dwavesys.com/docs/latest/doc_physical_properties.html}




\end{thebibliography}
\end{document}